\documentclass[10pt,conference]{IEEEtran}
\IEEEoverridecommandlockouts

\usepackage{multirow}
\usepackage{array}
\usepackage{siunitx}
\usepackage{cite}
\usepackage{amsmath,amssymb,amsfonts, amsthm}
\usepackage{tikz}
\usetikzlibrary{automata, positioning, arrows,calc}
\usetikzlibrary{quantikz}
\usepackage{algorithmic}
\usepackage{graphicx}
\usepackage{textcomp}
\usepackage{xcolor}
\usepackage{caption} 
\usepackage{subcaption}

\usepackage{hyperref}
\usepackage{mathtools}
\usepackage{multirow}
\usepackage{graphicx}
\usepackage{multicol}
\usepackage{array}
\usepackage{enumitem}
\usepackage{soul}

\newtheorem{cor*}{Corollary}

\newcommand{\mpfont}{\scriptsize}

\ifx\noeditingmarks\undefined
    \newcommand{\MPworker}[2]{{\color{#1}\vrule\vrule}{\marginpar{\color{#1}\mpfont #2}}}
\else
    \newcommand{\MPworker}[2]{}
\fi

\begin{document}

\title{Towards Quantum Networks:\\
Characterizing Raman Noise \\
over Metropolitan-scale Fiber Network

}

\makeatletter
\newcommand{\linebreakand}{%
  \end{@IEEEauthorhalign}
  \hfill\mbox{}\par
  \mbox{}\hfill\begin{@IEEEauthorhalign}
}
\makeatother

\author{
\IEEEauthorblockN{Marcello Caleffi~\IEEEmembership{Senior~Member,~IEEE} , Laura d'Avossa ~\IEEEmembership{Graduate Student~Member,~IEEE}, \\
Italo Ignacio Machuca Flores, Marco Grillo, Elena Montella, Angela Sara Cacciapuoti ~\IEEEmembership{Senior~Member,~IEEE}  }

\IEEEauthorblockN{www.QuantumInternet.it Research Group, University of Naples Federico II, Naples, Italy}

\\

\textsc{Invited Paper}

\thanks{
This work has been funded by the European Union under Horizon Europe ERC-CoG grant QNattyNet, n.101169850. Views and opinions expressed are however those of the author(s) only and do not necessarily reflect those of the European Union or the European Research Council Executive Agency. Neither the European Union nor the granting authority can be held responsible for them.}

}

\maketitle

\begin{abstract}
The coexistence of quantum and classical signals in optical fiber infrastructures represents a major challenge for large-scale quantum networks, as noise sources such as Raman scattering can significantly impact entanglement distribution, and the quantum protocols based on it. In this work, we analyze Raman scattering in the C-band, used for entanglement distribution, generated by a classical the O-band signal. The main contribution of this study lays in investigating these effects in a real metropolitan-fiber network, moving beyond controlled laboratory experiments to deployed telecommunication environments. Measurements are performed over a 7 km metropolitan fiber link using commercial sources and narrowband lasers.
The experimental results shows good agreement between the measurements taken under laboratory conditions, although, within the metropolitan-scale loop, localized spectral anomalies are observed in the deployed fibers.
Therefore, our results show that, whenever a quantum signal propagates in the C-band alongside an O-band classical channel within the same fiber, careful selection of the operating frequency is required, as Raman scattering and other real-world noise sources can significantly affect the quality and stability of the quantum transmission. In particular, we identify spectral regions that are less affected by Raman noise, thereby providing practical guidelines for optimal quantum channel allocation.
We demonstrate that Raman-induced noise constitutes a dominant contribution to the quantum signal-to-noise ratio (SNR) in realistic deployments, beyond background and detector noise. Overall, our findings offer practical insights for deploying quantum communication systems over existing fiber networks, supporting the development of robust and scalable quantum infrastructures.
\end{abstract}


\section{Introduction}
\label{sec:1}
Optical photons are considered ideal carriers of entanglement due to their physical properties. Indeed, they weakly interact with the environment, they have low transmission losses, and they can be precisely manipulated using standard optical components. These characteristics make them particularly well-suited for distributing entanglement between distant nodes in a quantum network \cite{CacCalTaf-19, CalDavHan-25, DavCacCal-25}.
\begin{figure*}[t]
    \centering
    \includegraphics[width=0.95\textwidth]{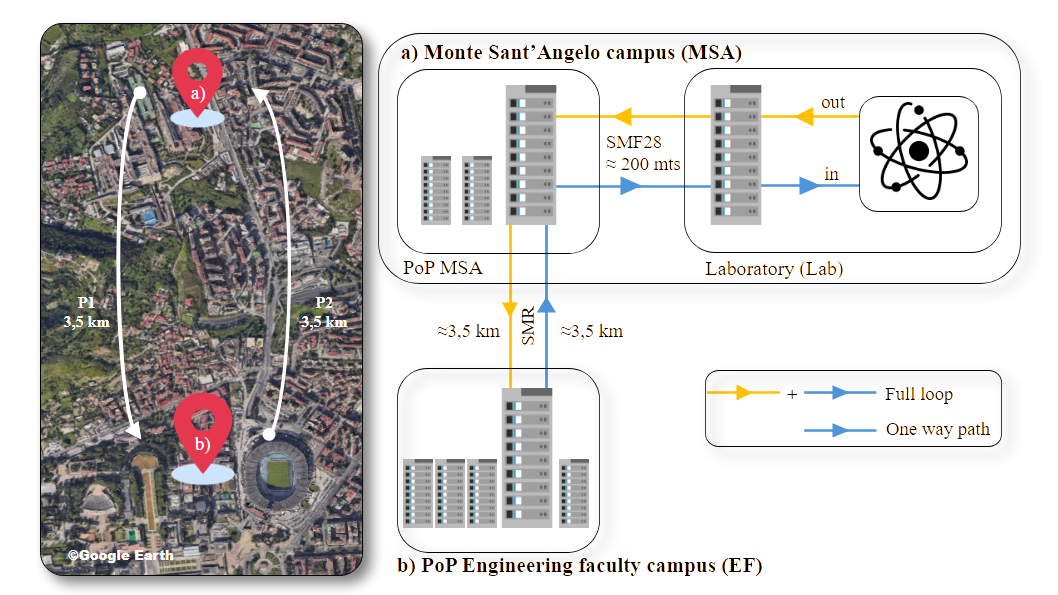} 
    \caption{\textbf{Scheme of the 7Km urban fiber loop of University of Naples Federico II.} The first path, P1, connects Monte Sant’Angelo (MSA) campus to the Engineering Faculty (EF) campus. The loop is closed through the second path, P2, which enables the return transmission of photons back to MSA campus. a) MSA campus hosts the laboratory (Lab), where photons are generated and detected after completing the loop. The campus also contains the Point of Presence (PoP), which provides access to the metropolitan fiber network. b) At the Engineering Faculty (EF) campus, the branches P1 and P2 are interconnected through a patch fiber, thereby establishing the return path toward MSA campus.}
    \hrulefill

    \label{fig:01}
\end{figure*}

\begin{figure}[t]
    \centering

    \includegraphics[width=0.95\columnwidth]{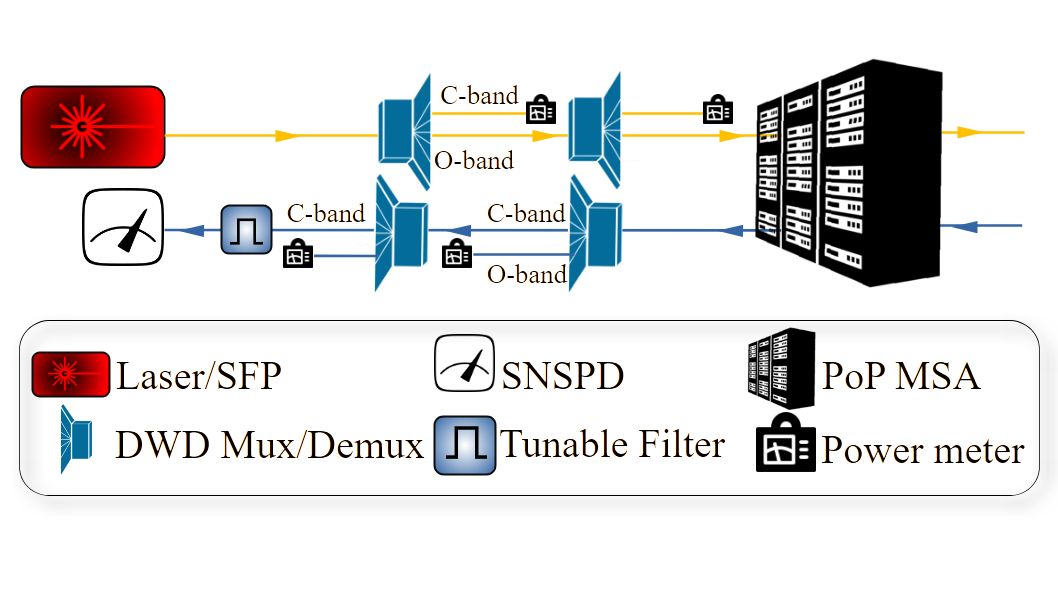} 
    \caption{\textbf{Experimental setup within the \textit{National QuantumInternet.it testbed}}. A classical signal in the O-band, generated either by a commercial SFP transmitter or by a narrow-linewidth laser source, is filtered through cascaded WDM modules and launched into the 7Km urban fiber loop. Raman-scattered photons generated in the C-band are spectrally selected using a tunable narrowband filter and detected by a superconducting nanowire single-photon detector (SNSPD).}
    \hrulefill

    \label{fig:02}
\end{figure}
Preliminary demonstrations of entanglement distribution have been successfully carried out using dedicated fiber infrastructures \cite{KruMerAch-19}. However, practical deployment of urban quantum networks requires leveraging existing telecommunication fiber systems, as deploying an infrastructure exclusively dedicated to quantum traffic would be impractical and cost-prohibitive. In this context, one of the main challenges arises from the coexistence of quantum signals together with high-power classical traffic within the same fiber. Such coexistence is particularly challenging, primarily due to the large power gap between classical and quantum signals — up to twelve orders of magnitude in our testbed — as well as the presence of multiple noise sources. One of the dominant contributions originates from Raman scattering that falls within the quantum channel band \cite{DavMonGri-26, EraWal-10, FroDynLuc-15, HolCan-02, FerXavTemVon-14}.
Spontaneous Raman scattering (SpRS) is a primary mechanism by which photons propagating within a given spectral region are scattered into different wavelength bands. 
Consequently, when distributing entanglement over an already deployed fiber infrastructure, the co-propagation photons of classical traffic may be scattered into the spectral band of the quantum signal, potentially introducing additional photons that we can consider as ground-noise, thus reducing the fidelity of the entangled state.
Assessing the impact of Raman scattering is therefore crucial for identifying spectral regions less affected by noise and select optimal channels for entanglement distribution coexisting with classical  
traffic. Previous studies have already investigated the impact of Raman noise in scenarios involving the coexistence of classical and quantum signals. In particular, such effects have been analyzed in multi-core fiber systems \cite{WuRbDiS-25}, where spatial separation between channels is exploited to mitigate inter-core nonlinear interactions, as well as in time-interleaved transmission schemes \cite{WanRolBri-24}, where classical and quantum signals are multiplexed in time to minimize the temporal overlap between Raman-scattered noise and quantum detection windows.

In contrast to these approaches, in this work we consider a simultaneous co-propagation scenario within the same single-core standard single-mode fiber. 
This allows us to study Raman-induced noise under realistic field conditions without relying on spatial or temporal multiplexing 
techniques, where classical and quantum channels coexist in the same physical transmission medium.
The conducted Raman noise characterization is performed over a single-mode fiber loop of \textit{National Quantum Internet.it testbed}, within the metropolitan-scale inter-campus network of University of Naples Federico II, as detailed in Sec.\ref{sec:02}.
This configuration enables the characterization of Raman scattering under real operating conditions, including the effects of deployed fiber imperfections.
Indeed, although Raman scattering has been extensively studied in controlled laboratory environments \cite{DavMonGri-26, ThoKanKum-25}, this work aims to extend the investigation to real fiber optic infrastructures, using an urban network environment with real commercial light sources. Indeed, Raman behaviour in such realistic scenarios remains less explored and can differ significantly due to environmental and structural inhomogeneities.
Moreover, to the best of the authors’ knowledge, this work represents the first experimental demonstration of quantum networking, where the quantum signal is allocated to the C-band, whereas classical propagation occurs in the O-band, as opposed to the previously investigated reverse configuration \cite{TalHesDav-26, TalHesTho-26}.
The rationale for this choice is that the C-band exhibits the lowest attenuation among standard telecom fiber bands, being the most suitable for quantum communications.


\section{Experimental setup}
\label{sec:02}
In this study, we measure Raman scattering with different classical sources. Firstly, we used a commercial SFP transmitter\footnote{An SFP transmitter is a compact optical module that converts electrical signals into optical signals for transmission over optical fiber. In this experiment, the device used is a Finisar FTLF1321P1BTL module.} with emission power of approximately $-0.8$dBm.
The SFP module has a central wavelength of 1310 nm, which is commonly used in short-to-medium range classical optical communications, reducing signal distortion during propagation. However, commercial SFP modules can exhibit a relatively broad optical spectrum, not strictly limited to a narrow line-width source as is the case with conventional CW lasers used in the laboratory. Yet, our choice allows us to investigate conditions closer to a real-world implementation scenario, where commercial devices are used. Furthermore, to verify the results obtained, a narrow-band laser source at 1310 nm with an output power of approximately $-4.6$ dBm is also used. This approach allows us to compare the Raman noise generated by narrow linewidth sources with that induced by real commercial transmitters.

To manage and control the effects of the classical sources, we use a series of optical components, including Wavelength Division Multiplexer (WDM) and filters, to shape the spectral properties of the signals under study.
Specifically, the classical O-band signal is first filtered through a cascade of two WDM modules to remove any components in other bands  that could affect the measurements. At the destination, another cascade of WDM filters is applied to isolate the C-band signal, i.e. the Raman-scattered signal.

The classical signals are transmitted over a loop within the metropolitan-scale fiber network of University of Naples Federico II.
The loop connects two university campuses: Monte Sant'Angelo (MSA), where the optical signal is generated and measured, and a remote site located approximately 3.5 km, at the Engineering Faculty (EF) campus. The link is configured in a loop-back architecture, resulting in a total propagation distance of approximately 7 km (see Fig.~\ref{fig:01}). 

The C-band photons generated by Raman scattering from the O-band are filtered using a high-resolution tunable narrowband filter with a bandwidth of 25 GHz ($200$ $pm$) capable of wavelength tuning across the 1525–1565 nm range. The filter is tuned in wavelength steps of 500 pm to scan different spectral channels. This choice not only ensures a selective extraction of the desired spectral components, but also allows scanning below the channel bandwidth, allowing access to spectral regions within the DWDM channel\footnote{The portion of spectrum between $1520$ and $1577$nm within C-band is divided into 40 DWDM channels, each with a spectral spacing of 100 GHz.} .


This enables an accurate wavelength scan in the allowed range, allowing us to analyze how Raman scattering varies for each specific DWDM channel, and thus alloiwng us to identifiy the C-band channel least affected by noise and more suitable for the co-propagation of quantum and classical traffic.

The photons generated in the C-band are then measured using a superconducting nanowire single-photon detector (SNSPD). 
The detector bias current is tuned to achieve system detection efficiency (SDE) \cite{NatTanHad-12} around $80\%$, with a dark count rate of $100$ $cps$.
The high performance of this experimental setup enables an in-depth analysis of the Raman spectra, allowing for the identification of even the subtlest variations in noise across the different wavelength channels and providing precise information on the optimal channels for the co-propagation of quantum and classical signals.
Fig.~\ref{fig:02} shows the complete experimental setup installed at the \textit{National Quantum Internet.it testbed}, including the source for transmitting the classical signal in the O-band and the detection system for receiving photons in the C-band

\begin{figure*}[t]
    \centering
    \begin{subfigure}[t]{0.49\textwidth}
        \centering
        \includegraphics[width=\linewidth]{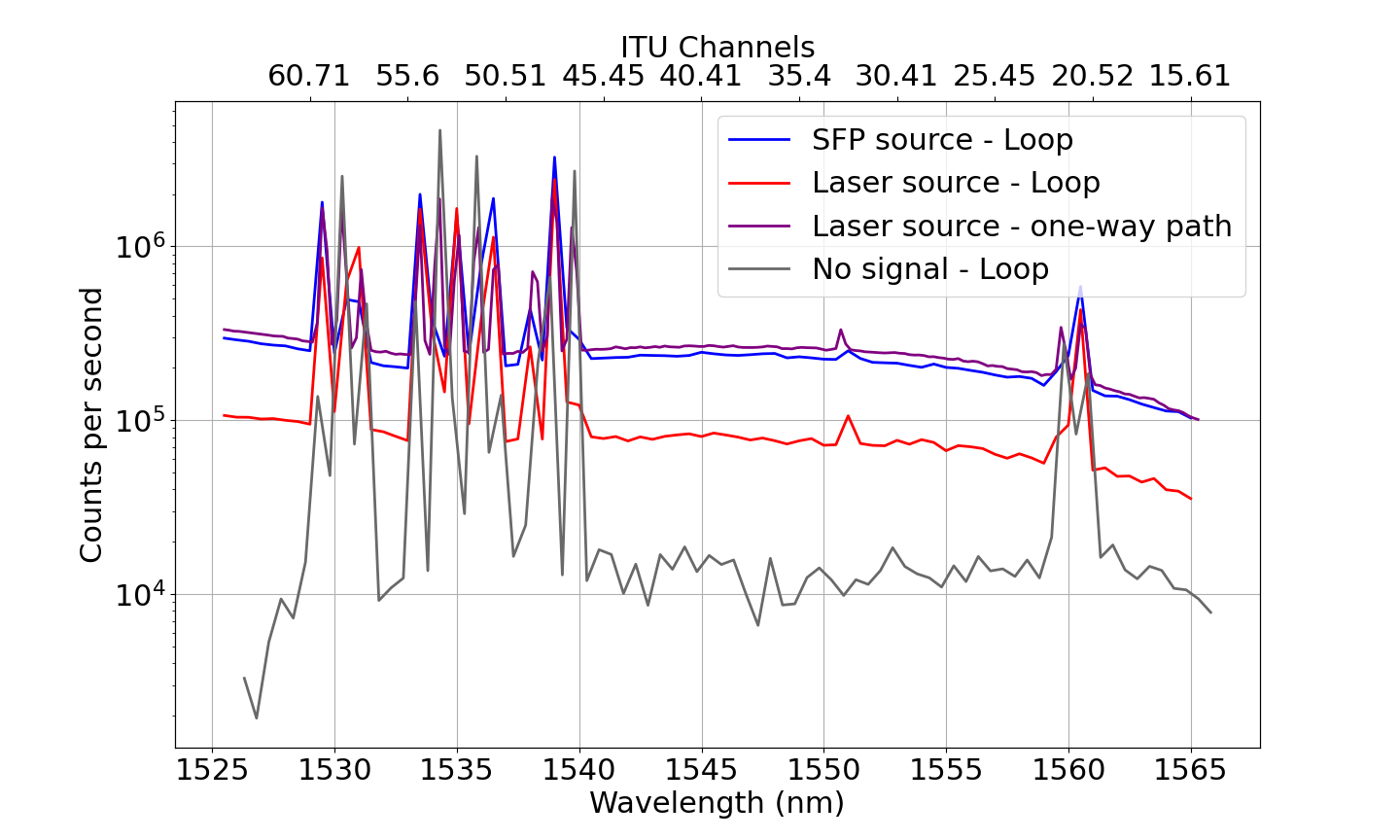}
        \caption{\textbf{Raw Data}}
        \label{fig:RawData}
    \end{subfigure}
    \hfill
    \begin{subfigure}[t]{0.49\textwidth}
        \centering
        \includegraphics[width=\linewidth]{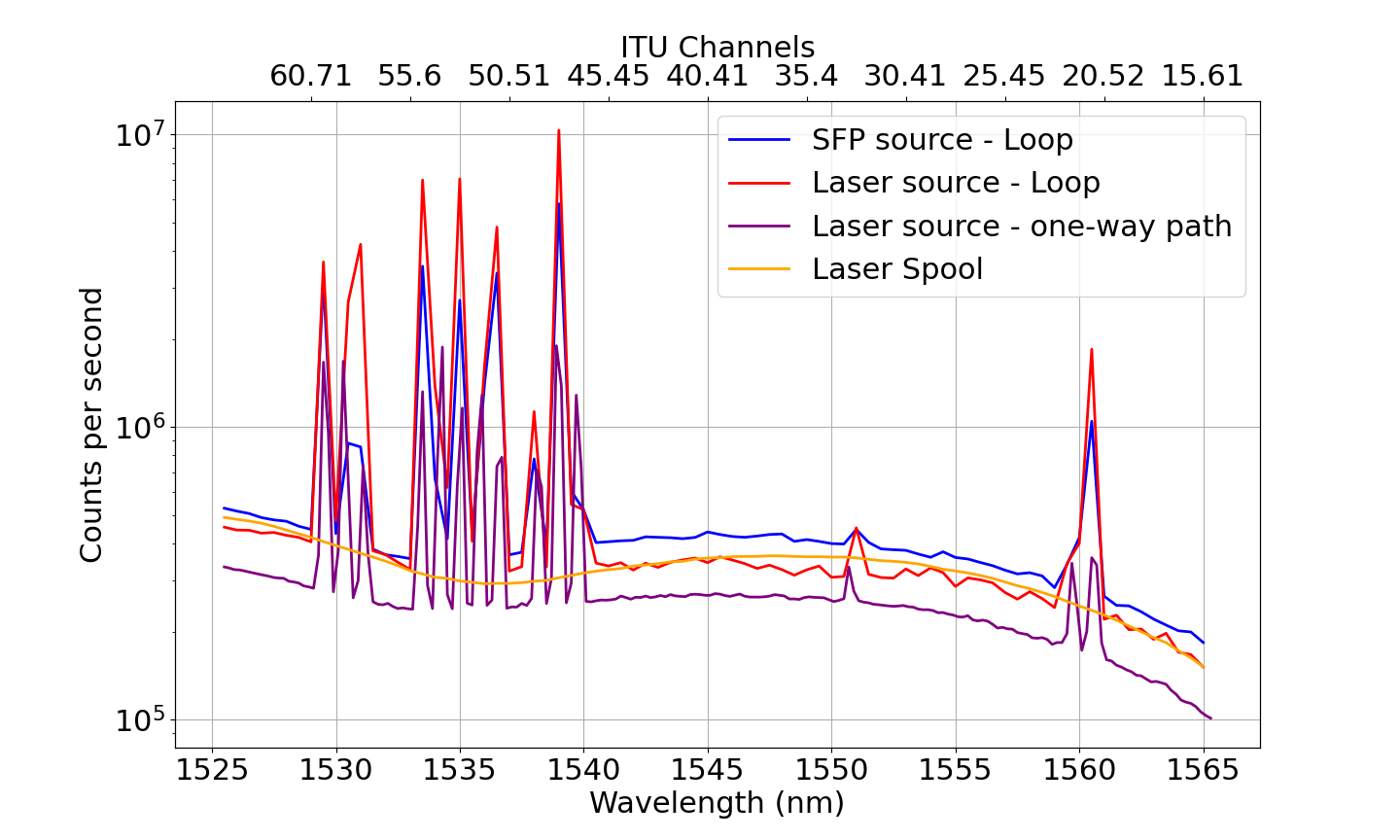}
        \caption{\textbf{Normalized Data}}
        \label{fig:NormalizedData}
    \end{subfigure}
    
    \caption{\textbf{Experimental results of Raman scattering spectra.} C-band SpRS photons generated from O-band classical sources are measured in terms of photon counts per second as a function of wavelength, corresponding to the ITU DWDM channels. The data include measurements obtained using both a narrowband laser source and a commercial SFP transmitter over the full fiber loop (7 km) and a one-way path (3.5 km). In (a) - Raw data, an additional reference measurement with no launched signal is included to identify background contributions. In (b) - Normalized data, an additional 5 km laboratory fiber spool measurement is reported, showing consistency between the urban-scale measurements and in-ab measurements after normalization.}
    \hrulefill

\end{figure*}

\section{Losses and Key Parameters}
A central objective of this study is to transfer results previously obtained under controlled laboratory conditions \cite{DavMonGri-26} to a deployed telecommunications network infrastructure (Details Fig. \ref{fig:01} and Fig.~\ref{fig:03}). As a first step, the attenuation along the loop was characterized through insertion-loss measurements performed on paths P1 and P2. Each path was divided into two contiguous fiber segments. The first segment connects the MSA laboratory to the PoP MSA using SMF-28 fiber, while the second segment extends from the same PoP to the PoP EF through an SMR fiber. The characterization procedure began by measuring the input optical power of a narrowband laser source, fixed at $-4.6$~dBm. The laser was sequentially tuned to operate at wavelengths of $1310$~nm and $1550$~nm, and the transmitted optical power was measured after each fiber segment. The insertion loss of each section was then determined from the difference between the input and output optical powers.

A detailed loss characterization of each optical segment is particularly relevant in inhomogeneous infrastructures, where are typically composed of different types of optical fibers and connectors. In the present implementation, the Lab-PoP MSA section employs SMF-28 fibers, with nominal attenuation coefficients of approximately $0.18$~dB/km at $1550$~nm and $0.32$~dB/km at $1310$~nm. By contrast, the PoP MSA - PoP EF section is based on SMR fibers, for which only the overall attenuation of the complete segment was available, measured to be $2.52$~dB at $1550$~nm. For all evaluated segments, the measured attenuation was lower at $1310$~nm. Moreover, as expected, the PoP MSA-PoP EF section in both paths, P1 and P2, exhibited the highest losses. Summary of the results is on Table \ref{tab:01}.

\begin{figure}[h!]
    \centering
    \includegraphics[width=0.5\columnwidth]{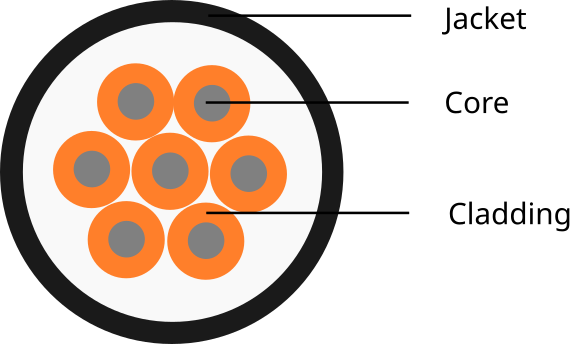} 
    \caption{\textbf{Transversal view of the metropolitan fiber}. The connection between the two campuses, MSA and EF, is established through a bundle of optical fibers. For our experiment, we use one of these channels, consisting of an SMF fiber with a core diameter of approximately $9~\mu$m and a cladding diameter of $125~\mu$m.}
    \hrulefill

    \label{fig:03}
\end{figure}


\begin{table}[t]
\centering
\renewcommand{\arraystretch}{1.8}
\caption{\textbf{Summary of the experiment losses.} Outputs power measured in each segments.}
	\hrulefill

\label{tab:01}

\scriptsize

\setlength{\tabcolsep}{9pt}

\begin{tabular}{|c|c|c|c|}
\hline
\textbf{Branch} & \textbf{Segment} & \textbf{1550 nm (dBm)} & \textbf{1310 nm (dBm)} \\
\noalign{\hrule height 1.3pt}

\multirow{2}{*}{P1} 
& Lab - PoP MSA & -3.1 dBm & -2.7 dBm \\
\cline{2-4}
& PoP MSA - PoP EF & -9.5 dBm & -3.1 dBm \\
\noalign{\hrule height 1.3pt}

\multirow{2}{*}{P2} 
& Lab - PoP MSA & -3.7 dBm & -3.1 dBm \\
\cline{2-4}
& PoP MSA - PoP EF & -8.5 dBm & -4.5 dBm \\
\hline

\end{tabular}
\hrulefill

\end{table}

\section{Results}
The experiment is implemented by measuring the C-band photons generated through Raman scattering by the two different O-band sources. This approach enables a direct comparison between a narrowband source and a commercially available SFP transmitter. Measurements are performed over the entire fiber loop, corresponding to the blue and red curves in Fig.~\ref{fig:01}.
To further investigate the Raman spectrum, an additional measurement with the narrowband source is carried out on half of the fiber loop (3.5 km), i.e. the purple curve in Fig.\ref{fig:01}. In this case, the narrowband source is used, with the signal launched from PoP EF and detected at MSA Lab.
The raw experimental results are reported in Fig.~\ref{fig:RawData}. From the measurements, a Raman spectral profile with a generally similar trend can be observed for the different configurations (both links and classical sources). However, distinct peaks appear at specific wavelengths, increasing the photon counts by up to an order of magnitude. To investigate the origin of these features, an additional measurement is performed with the loop in the absence of any launched signal. This measurement is also shown in Fig.~\ref{fig:RawData}, where peaks are observed at the same wavelengths as those detected in the previous measurements.
This indicates that the observed peaks are not associated with Raman-scattered signals, but rather originate from imperfections or damaged sections of the fiber within the loop.

The photon count measurements obtained under different experimental conditions are then normalized to ensure a consistent comparison. First, all datasets are scaled according to the optical power of the source, accounting for the different emission levels of the narrowband laser and the SFP module. Additionally, the measurements are normalized taking into account for different effective fiber lengths in each configuration, namely 7 km for the loop, 3.5 km for one way path of the loop. The resulting normalized spectra are presented in Fig.~\ref{fig:NormalizedData}.
Additionally, we validate the Raman profile measured in the urban fiber loop against a reference obtained in a noise-free laboratory environment. This reference, yellow line, measurement is performed using a narrowband laser over a 5 km fiber spool within the \textit{National Quantum Internet.it testbed} \cite{DavMonGri-26}. The recorded photon counts are rescaled to account for fiber attenuation (0.35 dB/km), enabling a direct comparison with the urban loop data. As shown in Fig.~\ref{fig:NormalizedData}, the two datasets exhibit strong agreement, confirming the consistency of Raman scattering behavior between deployed and laboratory fiber conditions.
The only noticeable differences are associated with the presence of spectral peaks, which, as previously demonstrated, are not related to the Raman signal. This provides further evidence that these features originate from imperfections in the deployed urban fiber.

\section{Discussion}

This work provides a comprehensive investigation of Raman noise by bridging the gap between controlled laboratory environments and real-world fiber infrastructures. By transitioning from laboratory conditions to an urban deployment scenario, we evaluate the behaviour of Raman scattering under realistic operating conditions, including the use of commercial optical sources and metropolitan fiber links.

The measurements performed in the urban loop are consistent with previous results obtained in controlled laboratory environments \cite{DavMonGri-26}. In particular, despite the transition from in-lab fiber spool to a real metropolitan fiber infrastructure, the overall Raman noise behaviour and spectral trend remain highly coherent with the laboratory observations. This close agreement demonstrates that the controlled-environment measurements accurately capture the dominant Raman scattering mechanisms and reliably predict the behaviour observed in practical field deployments.
Furthermore, these results provide further confirmation of the accuracy of the theorethical modelling developed from laboratory measurements, such as the one presented in \cite{DavMonGri-26}, demonstrating their ability to accurately describe Raman noise behaviour even in real urban scenarios.

However, the measurements performed in the deployed urban loop also reveal localized spectral anomalies that are not observable in controlled laboratory conditions.
The analysis of these anomalies is on-going and it is focusing on the physical structure of the deployed fiber loop, where multiple fibers are co-located within the same cable jacket (Fig. \ref{fig:03}). Such proximity can induce inter-fiber crosstalk effects, which may manifest as localized spectral features in the measured Raman noise. Indeed, although each fiber maintains its own core and cladding structure, their close physical proximity within the shared jacket can enable weak but non-negligible inter-fiber interactions, leading to the observed anomalies.

Although further work is needed to asses whether such spectral anomalies are effectively induced by the mentioned cross-talking, some phenomena seem to confirm this hypothesis. A representative example is the peak observed around 1560nm, corresponding to DWDM channels 21 and 22, namely, the same channels used for transmitting classical traffic within the University campus fiber loop. 

These findings motivate a revised definition of the Signal-to-Noise Ratio (SNR) for quantum communications systems operating in realistic deployment scenarios. While conventional noise sources—such as ambient light and detector dark counts—establish a baseline, particularly relevant for highly sensitive SNSPDs, they are not sufficient to fully describe practical operating conditions. In deployed fiber environments, Raman scattering generated by co-propagating classical signals introduces an additional and often dominant noise contribution. Therefore, an accurate SNR definition must explicitly account for both conventional background noise and Raman-induced noise to properly characterize system performance.

By identifying spectral regions where the overall SNR is optimized, our results provide practical guidelines for the selection of quantum channels and contribute to the development of more robust and reliable quantum communications systems in real-world deployments.

\bibliographystyle{IEEEtran}
\bibliography{bibliography.bib}

@article{CacCalTaf-19,
  title={Quantum internet: Networking challenges in distributed quantum computing},
  author={Cacciapuoti, Angela Sara and Caleffi, Marcello and Tafuri, Francesco and Cataliotti, Francesco Saverio and Gherardini, Stefano and Bianchi, Giuseppe},
  journal={IEEE Network},
  volume={34},
  number={1},
  pages={137--143},
  year={2019},
  publisher={IEEE}
}

@article{CalDavHan-25,
  title={Quantum transduction: Enabling quantum networking},
  author={Caleffi, Marcello and D’Avossa, Laura and Han, Xu and Cacciapuoti, Angela Sara},
  journal={IEEE Communications Surveys \& Tutorials},
  year={2025},
  publisher={IEEE}
}

@article{EraWal-10,
  title={Quantum key distribution and 1 Gbps data encryption over a single fibre},
  author={Eraerds, Patrick and Walenta, Nino and Legr{\'e}, Matthieu and Gisin, Nicolas and Zbinden, Hugo},
  journal={New Journal of Physics},
  volume={12},
  number={6},
  pages={063027},
  year={2010}
}

@article{HolCan-02,
  title={Multiple-vibrational-mode model for fiber-optic Raman gain spectrum and response function},
  author={Hollenbeck, Dawn and Cantrell, Cyrus D},
  journal={Journal of the Optical Society of America B},
  volume={19},
  number={12},
  pages={2886--2892},
  year={2002},
  publisher={Optical Society of America}
}

@inproceedings{ThoKanKum-25,
  title={Multiphoton interference and quantum teleportation coexisting with classical communications in optical fiber},
  author={Thomas, Jordan M and Kanter, Gregory S and Kumar, Prem},
  booktitle={Quantum Computing, Communication, and Simulation V},
  volume={13391},
  pages={98--112},
  year={2025},
  organization={SPIE}
}

@article{FroDynLuc-15,
  title={Quantum secured gigabit optical access networks},
  author={Fr{\"o}hlich, Bernd and Dynes, James F and Lucamarini, Marco and Sharpe, Andrew W and Tam, Simon W-B and Yuan, Zhiliang and Shields, Andrew J},
  journal={Scientific reports},
  volume={5},
  number={1},
  pages={18121},
  year={2015},
  publisher={Nature Publishing Group UK London}
}

@article{FerXavTemVon-14,
  title={Impact of Raman scattered noise from multiple telecom channels on fiber-optic quantum key distribution systems},
  author={Ferreira da Silva, Thiago and Xavier, Guilherme B and Tempor{\~a}o, Guilherme P and von der Weid, Jean Pierre},
  journal={Journal of lightwave technology},
  volume={32},
  number={13},
  pages={2332--2339},
  year={2014},
  publisher={OSA}
}

@article{KruMerAch-19,
  title={Light-matter entanglement over 50 km of optical fibre},
  author={Krutyanskiy, Viktor and Meraner, Martin and Schupp, Josef and Krcmarsky, Vojtech and Hainzer, Helene and Lanyon, Ben P},
  journal={npj Quantum Information},
  volume={5},
  number={1},
  pages={72},
  year={2019},
  publisher={Nature Publishing Group UK London}
}

@article{NatTanHad-12,
  title={Superconducting nanowire single-photon detectors: physics and applications},
  author={Natarajan, Chandra M and Tanner, Michael G and Hadfield, Robert H},
  journal={Superconductor science and technology},
  volume={25},
  number={6},
  pages={063001},
  year={2012},
  publisher={IOP publishing}
}

@article{WuRbDiS-25,
  title={Integration of quantum key distribution and high-throughput classical communications in field-deployed multi-core fibers},
  author={Wu, Qi and Ribezzo, Domenico and Di Sciullo, Giammarco and Cocchi, Sebastiano and Ann Shaji, Divya and Alves Zischler, Lucas and Luis, Ruben and Serena, Paolo and Lasagni, Chiara and Bononi, Alberto and others},
  journal={Light: Science \& Applications},
  volume={14},
  number={1},
  pages={274},
  year={2025},
  publisher={Nature Publishing Group UK London}
}

@article{WanRolBri-24,
  title={Time-interleaved C-band Co-propagation of quantum and classical channels},
  author={Wang, Jing and Rollick, Brian J and Jia, Zhensheng and Huberman, Bernardo A},
  journal={Journal of Lightwave Technology},
  volume={42},
  number={11},
  pages={4086--4095},
  year={2024},
  publisher={IEEE}
}

@article{DavMonGri-26,
  author = {Laura d'Avossa and
            Elena Montella and Marco Grillo and
            Angela Sara Cacciapuoti  and
            Marcello Caleffi},
  title = {Optimization of C-band quantum traffic coexisting with O-band classical traffic: preliminary results},
  journal = {IEEE MeditCom 2026: IEEE International Mediterranean Conference on Communications and Networking},
  year = {2026}
}

@article{TalHesDav-26,
  title={Synchronized distribution of quantum entanglement coexisting with high-rate, broadband classical optical communications over a real-world fiber link},
  author={Talcott, Gina M and Hess, Ahnnika I and d'Avossa, Laura and Kohlert, Scott J and Yeh, Fei I and Chen, Jim Hao and Mambretti, Joe J and Rambo, Tim M and Kanter, Gregory S and Thomas, Jordan M and others},
  journal={arXiv preprint arXiv:2602.00253},
  year={2026}
}

@article{DavCacCal-25,
  title={Modeling quantum transduction for multipartite entanglement distribution},
  author={d'Avossa, Laura and Cacciapuoti, Angela Sara and Caleffi, Marcello},
  journal={IEEE Transactions on Communications},
  volume={73},
  number={11},
  pages={11707--11721},
  year={2025},
  publisher={IEEE}
}

@inproceedings{TalHesTho-26,
  title={Synchronized Entanglement Distribution Across Deployed Fiber Coexisting with Fully-Loaded C-band Optical Communications},
  author={Talcott, Gina M and Hess, Ahnnika I and Thomas, Jordan M and d’Avossa, Laura and Kohlert, Scott J and Yeh, Fei I and Chen, Jim Hao and Mambretti, Joe J and Rambo, Tim M and Kanter, Gregory S and others},
  booktitle={Optical Fiber Communication Conference},
  pages={Tu2K--2},
  year={2026},
  organization={Optica Publishing Group}
}

\end{document}